\documentclass[12pt, a4paper]{article}
\usepackage[utf8]{inputenc}

\usepackage{amsmath,caption,bbm,amsfonts}
\usepackage{longtable,threeparttable,booktabs}
\usepackage[toc,page]{appendix}
\usepackage{natbib}
\usepackage{graphicx}

\usepackage[top=1.1in, bottom=1.1in, left=1.1in, right=1.1in]{geometry}

\usepackage{color,soul}
\usepackage{multirow}
\usepackage{algorithm}
\usepackage{algpseudocode}
\usepackage{subcaption}

\DeclareCaptionStyle{italic}[justification=centering]{labelfont={bf},textfont={it},labelsep=colon}

\usepackage{graphicx,psfrag,epsf,textcomp}
\usepackage{dsfont}
\usepackage{natbib,setspace}
\usepackage{url,xcolor} 
\usepackage{booktabs, bm, paralist,mathpazo,tikz,longtable,microtype}
\usepackage{enumitem}

\usepackage[pdftex,colorlinks=true]{hyperref}
\definecolor{darkblue}{rgb}{0,0,.6}
\hypersetup{citecolor=darkblue,linkcolor=darkblue,urlcolor=darkblue}

\newsavebox\CBox

\date{}
\usepackage{orcidlink}

\begin{document}

\def\spacingset#1{\renewcommand{\baselinestretch}{#1}\small\normalsize} \spacingset{1}

\title{\bf Depth-based reconstruction method for incomplete functional data}
\author{
  Antonio El\'ias Fernández\footnote{Corresponding author: aelias@uma.es} \ \orcidlink{0000-0002-3078-549X}  \\
  Department of Applied Mathematics \\
  OASYS group\\ Universidad de M\'alaga \\
  \\
  Ra\'ul Jim\'enez  \\
  Department of Statistics \\
  Universidad Carlos III de Madrid \\
  \\
   Han Lin Shang \orcidlink{0000-0003-1769-6430} \\
   Department of Actuarial Studies and Business Analytics \\
   Macquarie University
}
  
\maketitle



\bigskip

\begin{abstract}
The problem of estimating missing fragments of curves from a functional sample has been widely considered in the literature. However, a majority of the reconstruction methods rely on estimating the covariance matrix or the components of its eigendecomposition, a task that may be difficult. In particular, the accuracy of the estimation might be affected by the complexity of the covariance function and the poor availability of complete functional data. We introduce a non-parametric alternative based on a novel concept of depth for partially observed functional data. Our simulations point out that the available methods are unbeatable when the covariance function is stationary, and there is a large proportion of complete data. However, our approach was superior when considering non-stationary covariance functions or when the proportion of complete functions is scarce. Moreover, even in the most severe case of having all the functions incomplete, our method performs well meanwhile the competitors are unable. The methodology is illustrated with two real data sets: the Spanish daily temperatures observed in different weather stations and the age-specific mortality by prefectures in Japan.

\vspace{.2in}
\noindent \textit{Keywords}: Functional Data, Partially Observed Data, Reconstruction, Depth Measures.
\end{abstract}

\spacingset{1.5}

\section{Introduction}

Partially observed functional data (POFD) are becoming more recurrent, invalidating many of the existing methodologies \citep{ramsay2005, ferraty2006}. Diverse case studies motivate the development of statistical tools for these types of data. For example, in medical studies, many data sets are recorded through periodical check-ups, and patients who miss appointments or devices that fail to record may be typical sources of censoring. These situations may present in different types of monitoring, such as ambulatory blood pressure, the health status of human immunodeficiency virus tests (HIV), growth curves, and the evolution of lung function \citep{James2000, James2001, Delaigle2013, Kraus2015, Delaigle2016}. \cite{Sangalli2009, sangalli2014} also consider POFD from aneurysm studies where the source of censoring comes from a prior reconstruction of the sample and posterior processing to make the data comparable across subjects. In demography, it is common that age-specific mortality rates for older ages are not completely observed due to the decreasing number of survivors \citep{mortalityData} and this cohort is the focus of actuarial science studies \citep{DAmato2011}. Other examples involve electricity supply functions that may not be completely observed because suppliers and buyers typically agree on prices and quantities depending on the market conditions \citep{lieblReco2019, lieblSys2019}.

Prominent literature has tackled estimating missing parts of partially observed functional data, providing several benchmark methods. Among them, the methods of \cite{Yao2005, Goldberg2014, Kraus2015, Delaigle2016, lieblReco2019}. In this article, a new method is presented and compared with two of the above benchmark methods. We selected those that provided the best performance to minimize the mean squared error on our simulations and considered case studies. The selected benchmark methods are the method of \cite{Kraus2015}, and the method of \cite{lieblReco2019}. Notably, \cite{Kraus2015} proposes a functional linear ridge regression model for completing functional data based on principal component analysis. \cite{Kraus2015} estimates the principal component scores of the incomplete functions from the completely observed functions and then uses a completion procedure for recovering the missing part of the functions by using the observed part. On the other hand, \cite{lieblReco2019} approach the problem by introducing an optimal linear regression operator based on a local linear kernel aiming to produce smoother results and trying to avoid artificial jumps between observed and reconstructed parts.

Our approach combines two novel functional tools: 
\begin{inparaenum}
\item[1)] a depth-based method for functional time series forecasting \citep{projection2021}; and
\item[2)] a functional depth for partially observed functional data \citep{POIFD2021}.
\end{inparaenum}

The new method is conceived for:
\begin{inparaenum}
 \item[1)] Scenarios where the estimation of the covariance function is complex or impossible. We remark that the available reconstruction methods strongly depend on a proper estimation of the covariance. Its estimators might be very sensitive and become unreliable for many analyses, in particular for principal component analysis \citep{Hubert2005}.  In addition, a low number of complete functions in the sample also hampers the estimation procedures and makes it impossible if all the sample functions are partially observed \citep{lieblReco2019}.
 \item[2)] Reconstructing non-smooth functions. Some methods are designed to deal with smooth and unaligned functions, and, in consequence, their results are smooth and aligned functions \citep{lieblReco2019}. Our goal is to provide a method that produces reconstructed functions remaining as accurate as possible in roughness and variability.
\item[3)] Adding interpretability. It might be useful to get a precise estimation and some insights into the final reconstruction drivers. 
\end{inparaenum}
We consider simulated and real data for illustrating the issues listed above. On the one hand, we consider yearly curves of Spanish daily temperatures. This data set is gathered at different weather stations spread along with the Spanish territory. On the other hand, we consider age-specific yearly mortality rates recorded at each Japanese prefecture (political territory division). 

The structure of this paper is as follows: Section~\ref{sec:method} introduces notation and the method. Section~\ref{sec:results} shows a variety of simulated results based on Gaussian Processes under different simulated regimes of partial observability. In addition, we illustrate the method's performance with a Spanish daily temperatures data set and the Japanese age-specific mortality rates by prefecture. In Section~\ref{sec:conclu}, we make some conclusions.

\section{Depth-based reconstruction method}\label{sec:method}

\subsection{Definition and notation}

Let $X = \{X(t): t\in [a,b]\}$ be a stochastic process of continuous trajectories and $(X_1, \dots, X_n)$ independent copies of $X$. To simplify the notation, we assume without loss of generality $[a, b] = [0,1]$. We consider the case $X_1, \dots, X_n$ are partially observed. Following \cite{Delaigle2013}, we model the partially observed setting by considering a random mechanism $Q$ that generates compact subsets of $[0,1]$ where the functional data are observed. Specifically, let $O$ be a random compact set generated by $Q,$ and let $(O_1, \dots, O_n)$ be independent copies of $O$. Therefore, for $1\leq i \leq n,$ the functional datum $X_i$ is only observed on $O_i$. Let $ (X_i, O_i) = \{X_i(u): u \in O_i\}$, $M_i = [0,1]\setminus O_i$ and $(X_i, {M_i})$. Then, the observed and missing parts of $X_i$ are $(X_i, {O_i})$ and $(X_i, {M_i})$. We assume that $(X_1,O_1), \dots, (X_n,O_n)$ are i.i.d. realizations from $P \times Q$. This is, $\{X_1, \dots, X_n\}$ and $\{O_1, \dots, O_n\}$ are independent samples. This assumption, termed Missing-Completely-at-Random, is standard in the literature of POFD. Notedly, \cite{lieblSys2019} has considered a specific violation of the Missing-Completely-at-Random assumption.

The core idea of our method is based on the depth-based method for dynamic updating on functional time series \citep{projection2021}. In this context, but using the notation introduced here, the functional sample, ($X_1, \dots, X_n$), was ordered in time, $n$ being the most recent time period. $X_n$ was only observed on $O_n = [0,q]$, with $0\ll q< 1$, and the rest of sample curves $\{X_j:j< n\}$ were fully observed on $[0,1]$. This is, for all $j<n$, $O_j = [0,1]$. We may summarize the depth-based approach for dynamic updating as follows: if $(X_n,O_n)$ is depth in $\{(X_j,O_n) : j \in {\cal J}_n\}$, for some ${\cal J}_n\subset \{1,...,n-1\}$, and the band delimited by the curve segments $\{(X_j,O_n) : j \in {\cal J}_n\}$ captures both the shape and magnitude of $(X_n,{O_n})$, we may estimate $(X_n,M_n)$ from $\{(X_j,M_n): j\in {\cal J}_n\}$. In particular, point estimators of $(X_n,M_n)$ were obtained by computing weighted averages on the curve segments of $\{(X_j,M_n): j\in {\cal J}_n\}$. In our jargon, $\{(X_j,O_n): j\in {\cal J}_n\}$ is called the envelope of $(X_n,{O_n})$.

In contrast to the dynamic updating framework described above, the sample curves may not be temporarily ordered in the partially observed scenario that we are considering. What is more important, every single curve may be partially observed. So, for enveloping the observed part of a curve, say us $(X_i, {O_i})$, we could only have curve segments, namely $\{(X_j,O_i\cap O_j) : j \neq i\}$. Similarly, we may only have curve segments, specifically $\{(X_j,M_i\cap O_j) : j \neq i\}$, for estimating the missing part of $X_i$, that is $(X_i,M_i)$. How to apply the depth-based approach for these cases is a challenging problem that we address here. The method can be explained in two steps: one concerned about how to compute an \emph{envelope} of $(X_i,O_i)$, described in Section~\ref{subsec:curvesSelection} and one on how to estimate or reconstruct $(X_i,M_i)$ from the observed parts of the curves used for enveloping $(X_i,O_i)$, described in Section~\ref{subsec:reconstruction}.

\subsection{A depth-based algorithm for focal-curve enveloping}\label{subsec:curvesSelection}

This concept of depth arises for ordering multivariate data from \emph{the center to outward} \citep{liu1990, Liu1999, Rousseeuw1999, Zuo2000, Serfling2006, Li2012}. Let $ \mathcal{F}$ be the collection of all probability distribution functions on $\mathbb{R} $, $F\in \mathcal{F}$ and $x\in \mathbb{R}$. In the univariate context, a depth measure is a function $D:\mathbb{R} \times \mathcal{F}\rightarrow [0,1]$ such that, for any fixed $F$, $D(x,F)$ reaches their maximum value at the median of $F$, this is at $x$ such that $F(x)=1/2$, and decreases to the extent that $x$ is farthest from the median. Examples of such univariate depth measures are
\begin{equation}
\label{eq:univariateFM}
D(x,F) = 1- \Big| \frac{1}{2} - F(x)\Big|
\end{equation}
and 
\begin{equation}
\label{eq:univariateMBD}
D(x,F) = 2\big(F(x)(1-F(x))\big).
\end{equation}

Denote by $P$ to the generating law of the process $X$ and by $P_t$ to the marginal distribution of $X(t)$, this is $P_t(x) = \mathbb{P} ( X(t)\leq x)$. Given a univariate depth measure $D$, the Integrated Functional Depth of $X$ with respect to $P$ is defined as
\begin{equation}
\label{eq:fd}
\mbox{IFD}(X,P) = \int_0^1 D(X(t), P_t) w(t)dt,
\end{equation}
$w$ being a weight function that integrates to one \citep{Claeskens2014}. It is worth mentioning that when $w\equiv 1$, and $D$ is defined by Equation~\eqref{eq:univariateFM}, the integrated functional depth corresponds to the seminal Fraiman and Mu\~niz functional depth \citep{Fraiman2001}. When $D$ is defined by Equation~\eqref{eq:univariateMBD}, the corresponding IFD is the famous Modified Band Depth with bands formed by two curves \citep{lopezpintado2010}.

Define ${\cal J}(t) =  \{1\leq j \leq n: t\in O_j \}$. Suppose ${\cal J}(t) \neq \emptyset$ and let $q(t)$ be the cardinality of  ${\cal J}(t)$. Denote by $F_{{\cal J}(t)}$ to the empirical distribution function of the univariate sample $ \{X_j(t): j\in {\cal J}(t)\}$. This is the probability distribution that assigns constant mass equals to $1/q(t)$ to each available observation at time $t$. 
Then, for any pair $(X_i,O_i)$ and a given univariate depth $D$, we consider the empirical Partially Observed Integrated Functional Depth restricted to $O_i$ \citep{POIFD2021} defined by
\begin{equation}
\label{eq:POIFD}
\mbox{POIFD}(X_i,O_i) =  \frac{\int_{O_i} D\Big(X(t), F_{{\cal J}(t)}\Big)   q(t)dt}{\int_{O_i} q(t) dt}.
\end{equation}

In line with the approach for dynamic updating introduced by \cite{projection2021}, we search for a set ${\cal J}$ of sample curves, as big as possible, such that $i\notin {\cal J}$ and:

\begin{asparaenum}
\item[1)] $(X_i,O_i)$ is deep in $ \{(X_j, O_j): j\in {\cal J} \cup \{i\}\}$, the deepest if possible. For measuring depth here we use the Partially Observed Integrated Functional Dept restricted to $O_i$ defined in~\eqref{eq:POIFD}.
\item[2)] $(X_i,O_i)$ is enveloped by $ \{(X_j,O_j): j\in {\cal J}\}$ as much as possible. Here, we say $(X_i,O_i)$ is \emph{more enveloped} by $ \{(X_j,O_j): j\in {\cal J}\}$ than by $ \{(X_j,O_j): j\in {\cal J}'\}$ if and only if
\begin{equation*}
\lambda\left( \left\{t\in O_i: \min_{j\in {\cal J}}X_j(t) \leq X_i(t) \leq \max_{j\in {\cal J}} X_j(t) \right\} \right)  > \lambda\left( \left\{t\in O_i: \min_{j\in {\cal J}'} X_j(t) \leq X_i(t) \leq \max_{j\in {\cal J}'} X_j(t) \right\} \right),
\end{equation*}
with $\lambda$ being the Lebesgue measure on $\mathbb{R}$.
\item[3)]  $\{(X_j,O_j): j\in {\cal J}\}$ contains near curves to $(X_i,O_i)$,  as many as possible. For measuring nearness, we use \emph{mean $L_2$ distance} between POFD. This is,
\begin{equation}
\rVert (X_i,O_i)-(X_j,O_j)\rVert  = \frac{\sqrt{\int_{O_i\cap O_j} |X_i(t)-X_j(t)|^2 dt}}{\lambda(O_i\cap O_j)}.
\end{equation}
\end{asparaenum}
 
Algorithm~\ref{alg:Alg1} provides a set of curves with the three features above that we call the \emph{$i$-curve envelope} and denote by $ {\cal J}_i$ from now on. The algorithm is a variation of Algorithm 1 of \cite{projection2021}, adapted to POFD. Algorithm~\ref{alg:Alg1} iteratively selects as many sample curves as possible, from the nearest to the farthest to $(X_i, O_i)$, for enveloping $(X_i, O_i)$ and increasing its depth.

\algdef{SE}[SUBALG]{Indent}{EndIndent}{}{\algorithmicend\ }%
\algtext*{Indent}
\algtext*{EndIndent}
\begin{algorithm}
\caption{Input: $i, \{(X_j, O_j): 1\leq j\leq n\}$. Output: $\mathcal{J}$}
\label{alg:Alg1}
\begin{algorithmic}[]
\small
\State \textbf{Initialize} $f = (X_i,O_i), {\cal Y} = \{(X_j,O_i\cap O_j): j\neq i\}$, ${\cal J} = \emptyset$ and $D(f|{\cal J},i)=0$
\While{ size of ${\cal Y} \geq 2 $ }
\State Let $y'$ be the nearest curve to $f$ from $ {\cal Y}$ and ${\cal N} = \{y'\}$
\For{ $y\in {\cal Y}\setminus \{y'\}$, from the nearest curve to the farthest from $f$,}
\State $j_y = \{j: (X_j,O_i\cap O_j) = y\}$
\State ${\cal J}^+=  {\cal J}\cup \{j: (X_j,O_i\cap O_j) \in  {\cal N}\}$
\If{$f$ is more enveloped by  ${\cal N}\cup \{y\}$ than by ${\cal N}$ or $O_{j_y}\setminus \Big( \cup_{j\in {\cal J}^+} O_j \Big)\neq \emptyset$}
\State ${\cal N} = {\cal N} \cup \{y\}$ 
\EndIf
\EndFor
\If{$D(f|{\cal J}\cup {\cal N},i) \geq D(f|{\cal J},i)$}
\State ${\cal J} = {\cal J} \cup  \{j: (X_j,O_i\cap O_j)\in {\cal N}\}$ 
\EndIf
\State ${\cal Y} = {\cal Y} \setminus  {\cal N}$
\EndWhile
\end{algorithmic}
\end{algorithm}

\begin{figure}[]
\begin{center}
\includegraphics[]{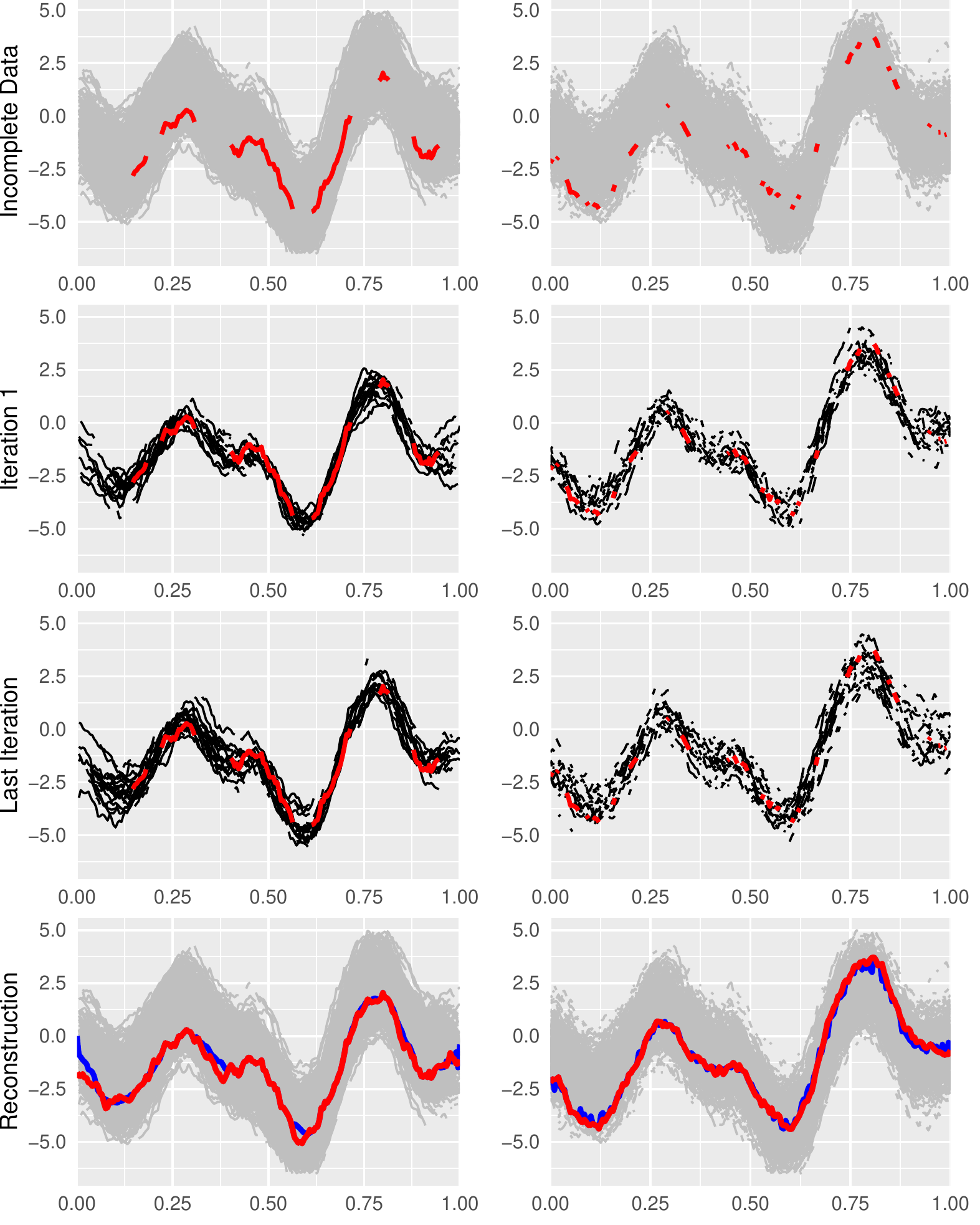}
\caption{\small{In the top panels, an illustration of how Algorithm~\ref{alg:Alg1} works. The sample curves correspond to $1000$ i.i.d. trajectories of a Gaussian process. We considered partially observed curves for the left panels by removing six random intervals from $[0,1]$. For the right panels, we considered missing data uniformly. On average, only 50\% of each curve was observed for both runs. The partially observed function that we reconstructed is colored in red and plotted entirely in the bottom panels jointly with its estimation. (in blue)}}
\label{fig:Algorithm}
\end{center}
\end{figure}

Figure~\ref{fig:Algorithm} illustrates how Algorithm~\ref{alg:Alg1} works. For there, we consider the first, second and final iteration of two runs from the algorithm based on $1000$ i.i.d. trajectories of a Gaussian process. We considered partially observed curves for the run shown in the left panels by removing six random intervals of the observation domain. For the run shown in the right panels, we considered missing data uniformly on the observation domain. On average, only 50\% of each curve were observed for both runs. The partially observed function we intend to reconstruct is colored in red and plotted entirely in the bottom panels jointly with its estimation that we describe how to compute below.

\subsection{Reconstruction of missing parts}\label{subsec:reconstruction}

For estimating the unobserved part of $X_i$, this is $(X_i, M_i)$, we use the same approach used for dynamic updating \citep{projection2021}. The estimator is a weighted functional mean from data of the curve envelope. Only here, these functional data may be partially observed. Specifically, these data are $\{(X_j,M_i\cap O_j) : j \in {\cal J}_i\}$. Consider ${\cal J}_i(t) = \{j\in {\cal J}_i:  t\in O_j\}$, assume ${\cal J}_i(t)\neq \emptyset$ for all $t\in O_i$, and let  $\delta = \min_{j\in {\cal J}_i} \rVert (X_i,O_i)-(X_j,O_j)\rVert $.
Then, we estimate $X_i$  on $M_i$ by
\begin{equation}\label{eq:pointestimator}
\widehat{X}_i^\theta (t)= \frac{\sum_{j\in {\cal J}_i(t)}  w_j  X_j(t)}{\sum_{j\in {\cal J}_i(t)}   w_j },\ \ \mbox{with} \ \ w_j = \exp\left(\frac{-\theta \rVert (X_i,O_i)-(X_j,O_j)\rVert}{\delta}\right).
\end{equation}
This estimator is a version of the envelope projection with exponential weights \citep[see][Equation (2)]{projection2021}, adapted to the partially observed data context. The parameter $\theta$ is chosen by minimizing the mean squared prediction error (MSE). In practice, if $\widehat{O}_i $ is the observational set where $\widehat{X}_i$ can be computable, this is $\cup_{j\in {\cal J}_i} (O_i\cap O_j)$, then $$\theta = \arg \min_{\nu}\sum_{i=1}^n\Vert (X_i, O_i)-(\widehat{X}^{\nu}_i,\widehat{O}_i)\rVert ^2.$$
As an illustration, the bottom panels of Figure~\ref{fig:Algorithm} show reconstructions of missing parts of the two simulated cases discussed above.

\section{Results}\label{sec:results}

We compare results obtained by using the depth-based method with those obtained from studies by \cite{Kraus2015} and \cite{lieblReco2019}. \cite{Kraus2015} propose a regularized regression model to predict the principal component scores (Reg. Regression) whereas \cite{lieblReco2019} introduce a new class of reconstruction operators that are optimal (Opt. Operator). The two methods were implemented by using the \textsf{R}-codes available at \url{https://is.muni.cz/www/david.kraus/web_files/papers/partial_fda_code.zip} and \url{https://github.com/lidom/ReconstPoFD}. 
The depth for partially observed curves and the data generation settings are implemented using the R-package \texttt{fdaPOIFD} of \cite{fdaPOIFD}.

Subsection~\ref{subset:simulation} introduces the simulation setting, the data generation process for POFD and shows results with synthetic data. Subsection~\ref{subsec:AEMET} uses the same simulation settings but applied to AEMET temperatures data. Additionally, it illustrates the reconstruction of some yearly temperature curves that are partially observed in reality. Finally, Subsection~\ref{subsec:JAPAN} presents another real case study where Japanese age-specific mortality functions are reconstructed.

\subsection{Simulation study}\label{subset:simulation}

Let us denote by $c\%$ the percentage of sample curves that are partially observed. We remark that benchmark methods perform better as the parameter $c$ is larger. This is because these reconstruction methods strongly depend on the information of the completely observed curves to estimate the covariance or the components of its eigendecomposition. However, the depth-based method can handle the case $c=0$, i.e., there are no complete functions in the sample. Therefore, results for this case are reported without comparison.

For our simulation study, we considered two Missing-Completely-at-Random procedures for generating partially observed data. These procedures have previously been used in the literature \citep{POIFD2021} and are in line with the partial observability of the real case studies. They are:
\begin{asparaitem}
\item[] \textbf{Random Intervals}, with which $c$\% of the sample curves is observed on a number $m$ 
of random disjointed intervals of $[0,1]$.
\item[] \textbf{Random points}, with which $c$\% of the functions is observed on a very sparse random grid.
\end{asparaitem}
First, we apply these observability patterns to simulated trajectories. Concretely, we consider a Gaussian process $X(t) = \mu (t) + \epsilon(t)$ where $\epsilon(t)$ is a centered Gaussian process with covariance kernel $\rho_{\epsilon}(s, t) = \alpha e^{-\beta | s - t |}$ for $s,t \in [0,1]$. The functional mean $\mu(t)$ is a periodic function randomly generated by a centered Gaussian process with covariance $ \rho_{\mu}(s,t) = \sigma e^{-(2\sin(\pi |s - t|)^2/l^2)}$.  Thus, each sample will present different functional means. The set of parameters used for our study were $\beta = 2$, $\alpha = 1$, $\sigma = 3$ and $l=0.5$.  Example of the generated trajectories by this model are those shown in Figure~\ref{fig:Algorithm}. 

We considered small and large sample sizes for the study by making $n= 200$ and $n = 1000$. Also, we considered different percentages of observed curves that were partially observed. Specifically, we tested with $c = 0, 25, 50$ and $75$. In addition, we considered different percentages of time on which the incomplete curves of a sample were observed. Henceforth, we term this percentage by $p\%$. We considered $p = 25, 50$ and $75$ for small samples but only $p = 25$ and $50$ for large samples. This is due to the computational cost of the benchmark methods when $n=1000$ and $p=75$. Note that this parameter setting implies the highest computational cost for estimating covariance functions. Finally, we replicate 100 samples of each data set for estimating median values of MSE.

\begin{figure}[!htbp]
\begin{center}
\includegraphics[width=18cm]{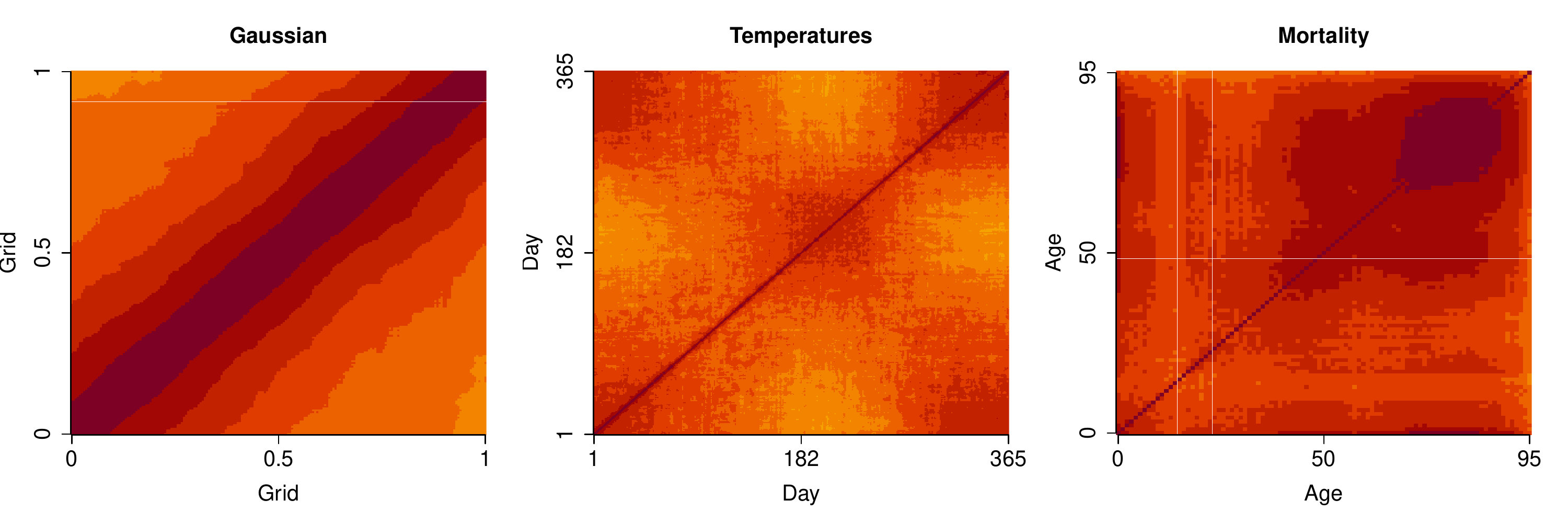}
\caption{\small{Covariance estimations based on the available functions are completely observed. Left panel, Gaussian processes with an exponential decay covariance.  Central panel: Spanish daily temperatures with lower covariance values in Spring and Autumn periods and higher covariance in Summer and Winter. Right panel: Japanese age-specific mortality rates with higher correlations at the oldest ages.}}
\label{fig:covariances}
\end{center}
\end{figure}

Table~\ref{tab:RandomPoints} presents results for $n=200$. It shows MSE from the Gaussian data and points out the superiority of the Reg. Regression method \citep{Kraus2015} when covariance function is simple to estimate, as is the case of the exponential decay covariance function involved in these data (see left panel of Figure~\ref{fig:covariances}). We remark that, even in this case, depth-based is lightly better than the Opt. Operator method \citep{lieblReco2019}. When all the functions of the sample are partially observed ($c=0\%$), only the depth-based method can provide a reconstruction and, surprisingly, the MSE remains reasonably similar to those cases with a proportion of complete functions significantly large ($c = 25, 50, 75 \%$). Similar results are obtained with larger sample sizes.

\begin{table}[!htbp]
\caption{\small{Median values of mean square errors over $100$ pseudo-random replicates. Each replicate is composed of $200$ curves. A dash (-) represents that the method cannot produce any reconstruction. The partially observed samples are obtained by observing $p\%$ of the total discrete realization points (Random Points). The smallest error is bolded for each combination of $c$ and $p$.}} \label{tab:RandomPoints}
\centering
\tabcolsep 0.05in
\resizebox{\textwidth}{!}{
\begin{tabular}{@{}lrrr|rrr|rrr|rrr@{}}
\toprule
& \multicolumn{3}{c}{$c=75$} & \multicolumn{3}{c}{$c=50$} &  \multicolumn{3}{c}{$c=25$} &  \multicolumn{3}{c}{$c=0$} \\ 
\cmidrule{2-13}
Method & $p=25$ & 50 & 75 &  25 & 50 & 75 & 25 & 50 & 75 & 25 & 50 & 75 \\
\midrule
Depth-based &  0.153 & 0.138 & 0.133  & 0.169 & 0.144 & 0.137 &  0.197 & 0.154 & 0.14 & \textbf{0.259} & \textbf{0.168} & \textbf{0.143} \\
Opt. Operator & 0.159 & 0.16 & 0.247  & 0.179 & 0.191 & 0.246 & 0.187 & 0.185 & 0.256 & - & - & - \\
Reg. Regression & \textbf{0.059} & \textbf{0.054} & \textbf{0.049} &  \textbf{0.075} & \textbf{0.07} & \textbf{0.06} &  \textbf{0.124} & \textbf{0.111} & \textbf{0.086}  & - & - & - \\
\bottomrule
\end{tabular}
}
\end{table}

\subsection{Case study: Reconstructing AEMET temperatures}\label{subsec:AEMET}

Spanish Agency of Meteorology (AEMET) provides meteorological variables recorded from different stations in the whole Spanish territory (see \url{http://www.aemet.es/es/portada}). This analysis focus on maximum daily temperatures of $73$ stations located in the capital of provinces. 
Following the literature of FDA, we consider this data as a functional data set where each function are the temperatures of each complete year \citep{fdausc, edu2014}. Some of the curves are partially observed in the historical data, and our goal is to reconstruct the data set. 

\vspace{.3in}

Temporal data availability depends from one station to the other. For example, Madrid-Retiro station is the oldest station, being monitored from $1893$. However, Ceuta from $2003$. In total, we consider a set of $2786$ entirely observed curves of different years and weather stations. This large sample of complete functions allows reproducing the simulation in Subsection~\ref{subset:simulation} by randomly generating random functional samples. To do that, we randomly generate $100$ samples of curves without replacement. The results for sample sizes of $n=200$ and $n=1000$ are in Table~\ref{tab:randomPointsBigsample}. Unlike Gaussian data, the depth-based method seems to be superior to the competitors in the simulation with AEMET data. Only for high proportions of complete functions $c=75$ and small sample size $n = 200$, \citeauthor{Kraus2015}'s \citeyearpar{Kraus2015} method was superior, becoming the depth-based method superior for smaller values of $c$ or larger sample sizes $n = 1000$. These results can be explained by the complex structure of AEMET data as shown in the center panel of Figure~\ref{fig:covariances}.

\begin{table}[!htbp]
\tabcolsep 0.13in
\caption{\small{Median values of mean square errors over $100$ replications. Each replicate is composed of $1000$ and $200$ curves (results between parenthesis). A dash (-) represents that the method cannot produce any reconstruction. The partially observed samples are obtained by observing $p\%$ of the total discrete realization points (Random Points).}} \label{tab:randomPointsBigsample}
\centering
\resizebox{\textwidth}{!}{%
\begin{tabular}{@{}lrr|rr|rr|rr@{}}
\toprule
 & \multicolumn{2}{c}{$c=75$}  & \multicolumn{2}{c}{$c=50$} &  \multicolumn{2}{c}{$c=25$} & \multicolumn{2}{c}{$c=0$} \\ 
 \cmidrule{2-9}
Method & $p=25$ & 50 &  25 & 50 &  25 & 50 & 25 & 50 \\
\midrule
Depth-based &  \textbf{5.393} & \textbf{4.920} &  \textbf{6.384} & \textbf{5.274}  & \textbf{7.874} & \textbf{6.321}  & \textbf{10.252} & \textbf{7.606}   \\
& (9.639) & (9.099) & (10.879) & \textbf{(9.585)} & \textbf{(12.555)} & \textbf{(10.079)} & \textbf{(14.691)} & \textbf{(10.900)} \\
Opt. Operator & 13.081 & 13.403 &  13.201 & 13.442 & 13.259 & 13.229 & - & - \\
& (13.184) & (13.530) & (13.296) & (13.493) & (13.412) & (13.436) & - & - \\
Reg. Regression & 6.834 & 5.41 & 7.349 & 5.944 & 8.617 & 8.217 & - & -  \\ 
& \textbf{(9.508)} & \textbf{(8.826)} & (10.842) & (10.472) & (13.577) & (13.251) & - & - \\
\bottomrule
\end{tabular}
}
\end{table}

Table~\ref{tab:AEMET_intervals} shows the same simulation setup with AEMET data but under the Random Interval setting and small sample size. In this setting, we generate partially observed data for a different number of observed intervals ($m$), percentages of completely observed curves ($p$), and mean observability percentage of each partially observed curve ($c$). The result shows that when $p=25$ the Reg. The regression method typically performs better than the competitors. This is because our implementation of the partially observed setting produces a sample of curves not as densely distributed in the extremes of the domain as it is in the middle of the domain. Then, for small $p$ to extrapolate using the depth-based method to dense part of the domain worsens the results. However, when $p$ increases, our implementation produces a sample of partially observed curves covering the complete domain densely, and the depth-based method outperforms.

\begin{table}[!htbp]
	\caption{\small{Median values of the mean squared error after $100$ replications. This table summarizes the exercise considering random samples from the fully observed AEMET data set. Each replicate is composed of $200$ functional observations. The partially observed samples are obtained by restricting each function to $m$ intervals of total length $p\%$ of the domain (Random Intervals).}}
	\label{tab:AEMET_intervals}
	\centering
	\resizebox{\textwidth}{!}{%
	\begin{tabular}{@{}llrrr|rrr|rrr@{}}
	\toprule
	& & \multicolumn{3}{c}{$c=75$} & \multicolumn{3}{c}{$c=50$} & \multicolumn{3}{c}{$c=25$}  \\ 
	\cmidrule{3-11}
Intervals & Method & $p=25$ & 50 & 75 &  25 & 50 & 75 & 25 & 50 & 75 \\
			\midrule
			$m = 1$ & Depth-based & \textbf{13.359} & \textbf{10.037} & \textbf{8.999} & 15.597 & \textbf{11.224} & \textbf{9.690} & 18.122 & \textbf{12.903} & \textbf{10.884} \\
			& Opt. Operator & 16.041 & 12.862 & 10.998 & 16.105 & 13.119 & 10.979 & 16.567 & 13.366 & 11.168  \\
			& Reg. Regression & 14.420 & 11.033 & 9.472 & \textbf{15.466} & 12.359 & 10.290 & \textbf{16.821} & 14.194 & 10.877 \\
			\addlinespace[0.5cm]
			$m = 2$ & Depth-based & 13.546 & \textbf{10.363} & \textbf{8.975} & 16.935 & \textbf{11.076} & \textbf{9.621} &  19.415 & \textbf{12.347} & \textbf{10.385} \\
			& Opt. Operator & 16.657 & 13.689 & 11.200 & 16.856 & 13.591 & 11.373 & 16.960 & 13.662 & 11.549 \\
			& Reg. Regression & \textbf{13.369} & 11.394 & 9.542 & \textbf{14.439} & 12.393 & 10.597 & \textbf{16.074} & 14.228 & 12.422  \\
			\addlinespace[0.5cm]
			$m = 4$ & Depth-based & 13.841 & \textbf{10.316} & \textbf{9.199} & 16.403 & \textbf{10.829} & \textbf{9.579} & 18.682 & \textbf{11.800} & \textbf{10.097} \\
			& Opt. Operator & 16.164 & 13.665 & 12.126 & 16.088 & 13.689 & 11.953 & 16.105 & 13.563 & 11.962  \\
			& Reg. Regression & \textbf{12.909} & 11.320 & 10.022 & \textbf{13.676} & 12.354 & 11.063 & \textbf{15.427} & 14.301 & 12.736 \\
			\hline 
		\end{tabular}%
	}
\end{table}

\begin{figure}[!htbp]
\begin{center}
\includegraphics[width=14cm]{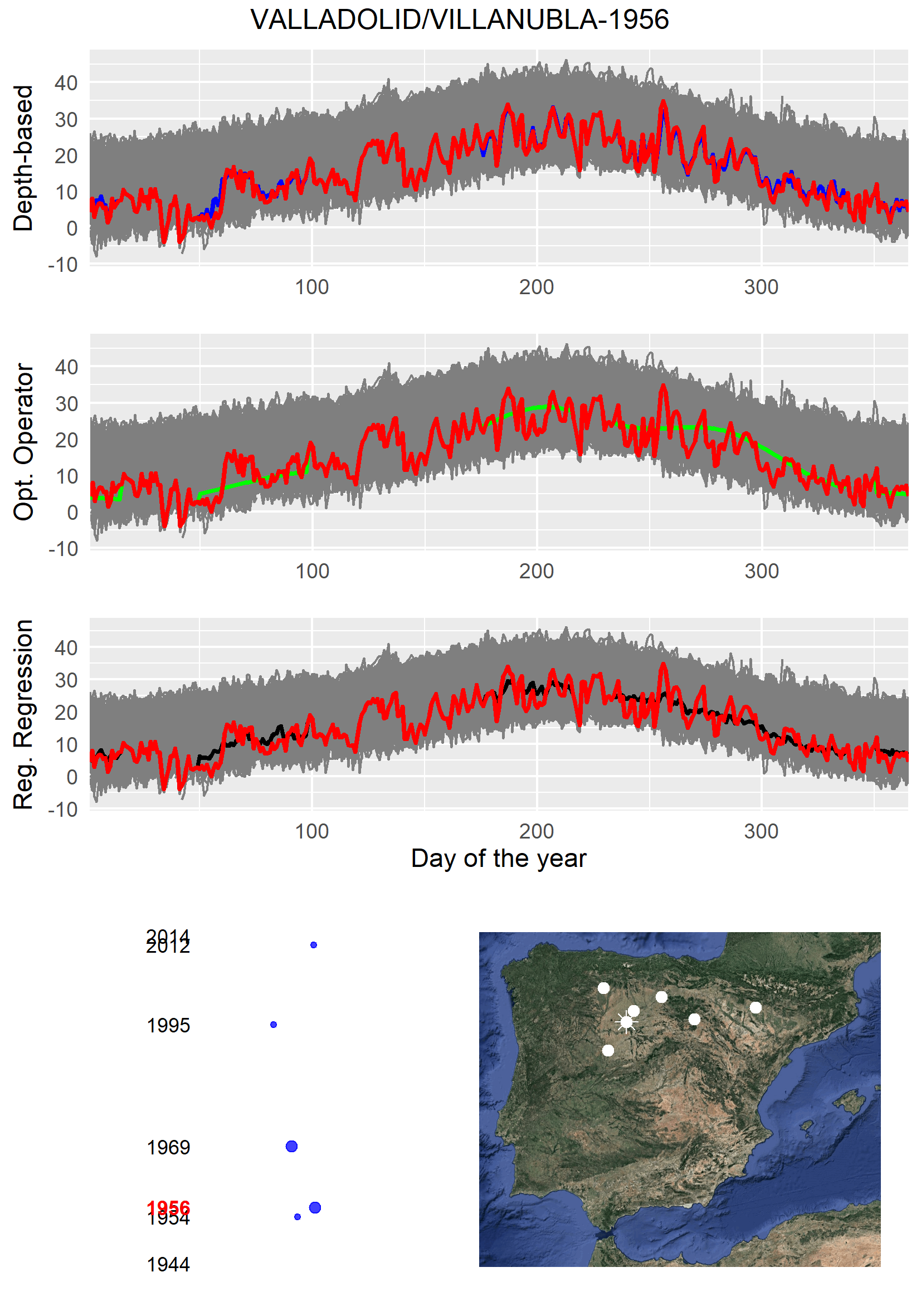}
\caption{\small{Simulated exercise reconstruction of ``VALLADOLID/VILLANUBLA-1956". Top panel presents the reconstruction by \cite{Kraus2015} (black) and \cite{lieblReco2019} (green). Middle panel, the reconstruction provided by the depth-based method. Bottom panel, the spatial (Spanish map) and temporal (bubble plot) descriptive analysis offered by the depth-based methodology and the envelope.}}
\label{fig:simulationAEMET}
\end{center}
\end{figure}

In Figure~\ref{fig:simulationAEMET}, we plot the reconstructions obtained by the three methods under consideration from one random sample. This was obtained by randomly taking $1000$ curves from the total observed curves of the AEMET data. Then, we generated partially observed data by applying the Missing-Completely-at-Random procedure based on random intervals described above, with $m = 4$, $p=50$, and $c=50$.
Finally, we randomly selected one function to reconstruct, namely, VALLADOLID/VILLANUBLA-1956, where VALLADOLID/VILLANUBLA refers to the location of the station and 1956 to the observation year. According to a general view of its shape, VALLADOLID/VILLANUBLA-1956 is completely plotted in red. In contrast, the reconstructions are only plotted on the four intervals where the curve was observed in our simulation. The top panels of the figure show output produced by the depth-based method (Depth-based in blue), \cite{lieblReco2019} (Opt. Operator in green), by \cite{Kraus2015} (Reg. Regression in black). The depth-based method is superior to the benchmark methods. The bottom panel of the figure shows some descriptive statistics related to the depth-based method. We show years used for reconstructing (the years of the curves into the envelope). The frequency of each year (number of curves into the envelope with the same year) is represented by a proportional blue bubble. Similarly, we show at the right side the locations of the curves into the envelope. From our understanding, all of them have similar geographical features.

\begin{figure}[!htbp]
\begin{center}
\includegraphics[height=14cm]{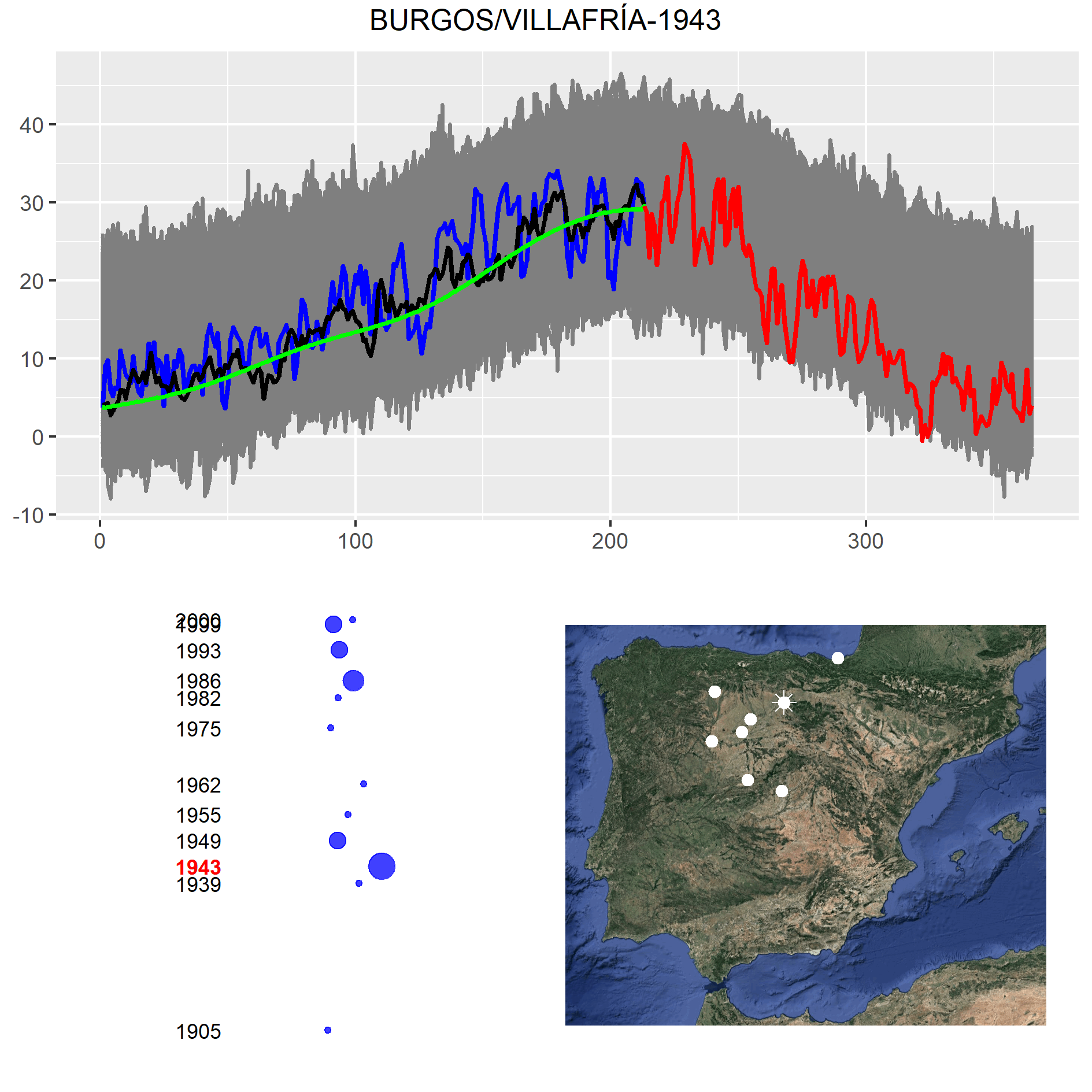}
\caption{\small{Real partially observed function, ``BURGOS/VILLAFRÍA-1943". Top panel: reconstructions by \cite{Kraus2015} (in black) \cite{lieblReco2019} (in green) and the depth-based method (in blue). The bottom panel shows the descriptive analysis of the most relevant curves in reconstructing ``BURGOS/VILLAFRÍA-1943". In the left part, time analysis (bubble plot of the involved years); in the right part, spatial analysis (map with the most relevant and involved stations).}}
\label{fig:realAEMET}
\end{center}
\end{figure}

Finally, Figure~\ref{fig:realAEMET} illustrates the actual case of ``BURGOS/VILLAFRÍA-1943" a station that probably started operating in the middle of the year $1943$. Consequently, only the year's second half is recorded (red curve at the top panel). We apply the three reconstructing methods to complete the first half of the curve (Reg. Regression \citep{Kraus2015} in black, Opt. Operator \citep{lieblReco2019} in green, and the depth-based method in blue). The depth-based methodology supports the reconstruction with the additional information provided by the most important curves of the envelope. The subsample contains distant-past function from 1905 from MADRID-RETIRO station, more recent functions from the $90$s from the same ``BURGOS-VILLAFRÍA" station a bigger proportion of functions from $1943$, belonging to the same year of the partially observed function to reconstruct. 

\subsection{Case study: Reconstructing Japanese mortality}\label{subsec:JAPAN}

The Human Mortality Data Set (\url{https://www.mortality.org}) provides detailed mortality and population data of $41$, mainly in developed countries. Some countries also offer micro information by subdivision of the territory, providing challenging spatial and temporal information. In particular, the Japanese mortality data set (\url{http://www.ipss.go.jp/p-toukei/JMD/index-en.asp}) is available for its $47$ prefectures for males, females, and the total population.

A common FDA approach to analyze mortality data is to consider that each function is the yearly mortality for each age cohort \citep{hanlinHyndman2017, ShangHaberman2018, gao2019, Shang2019}. With this configuration and arranging the $47$ prefectures together, we deal with a male, female, or total Japanese mortality data set of size $2007$. Of course, each prefecture does not have the same number of functions, and the range of observed years is also different. However, roughly, we have yearly mortality functions between 1975 and 2016.

\begin{figure}[!htbp]
\begin{center}
\includegraphics[width=14cm]{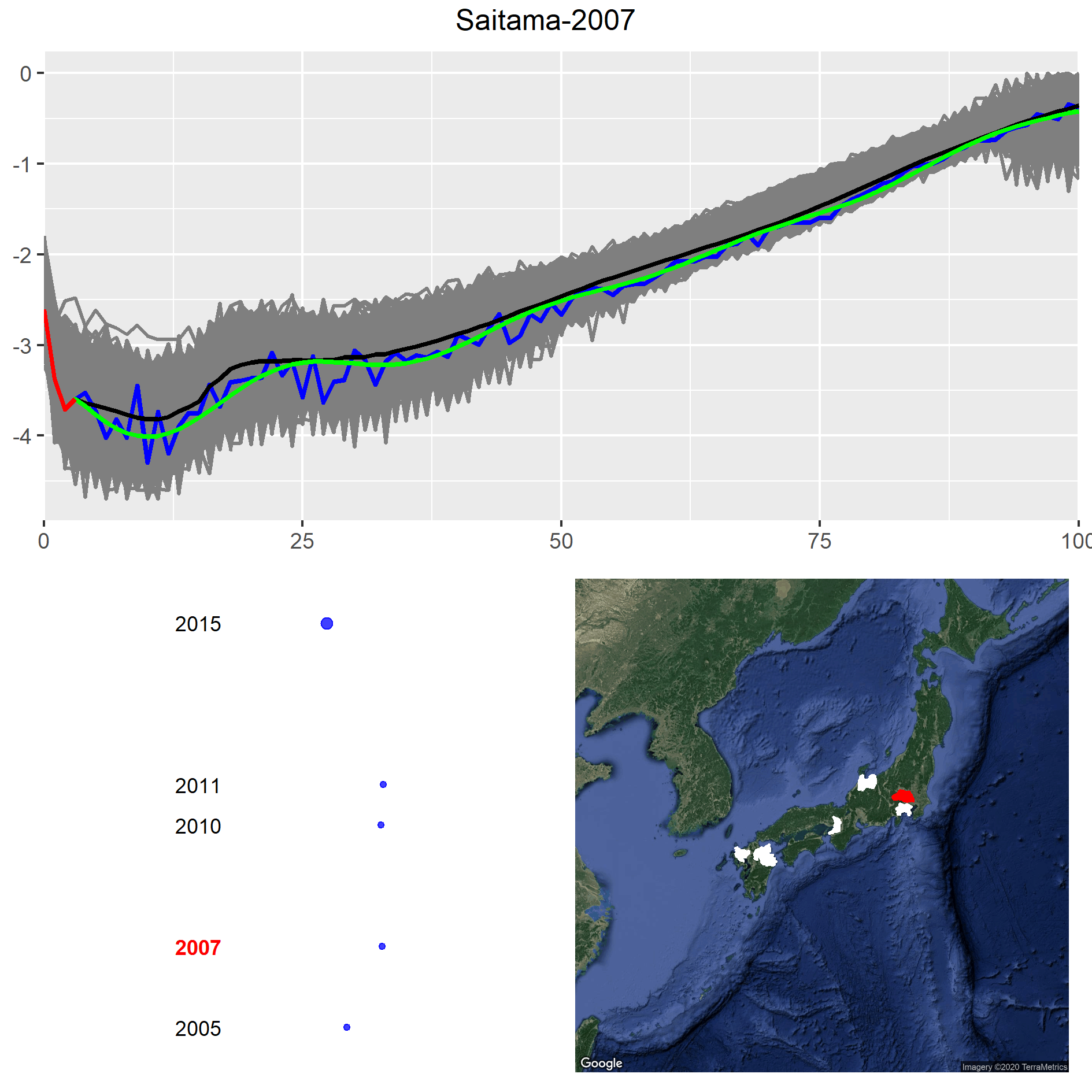}
\caption{\small{Reconstruction of the most poorly observed function of the sample, ``Saitama-2007". The top panel presents the reconstruction given by Reg. Regression method by \cite{Kraus2015} (black), Opt. Operator method by \cite{lieblReco2019} (green) and the depth-based method (blue). The bottom panel shows the year of the most important functions of the envelope, and the maps showing the corresponding prefectures, in red the one to be reconstructed.}}\label{fig:JAPANshortest}
\end{center}
\end{figure}

\begin{figure}[!htbp]
\begin{center}
\includegraphics[height=14cm]{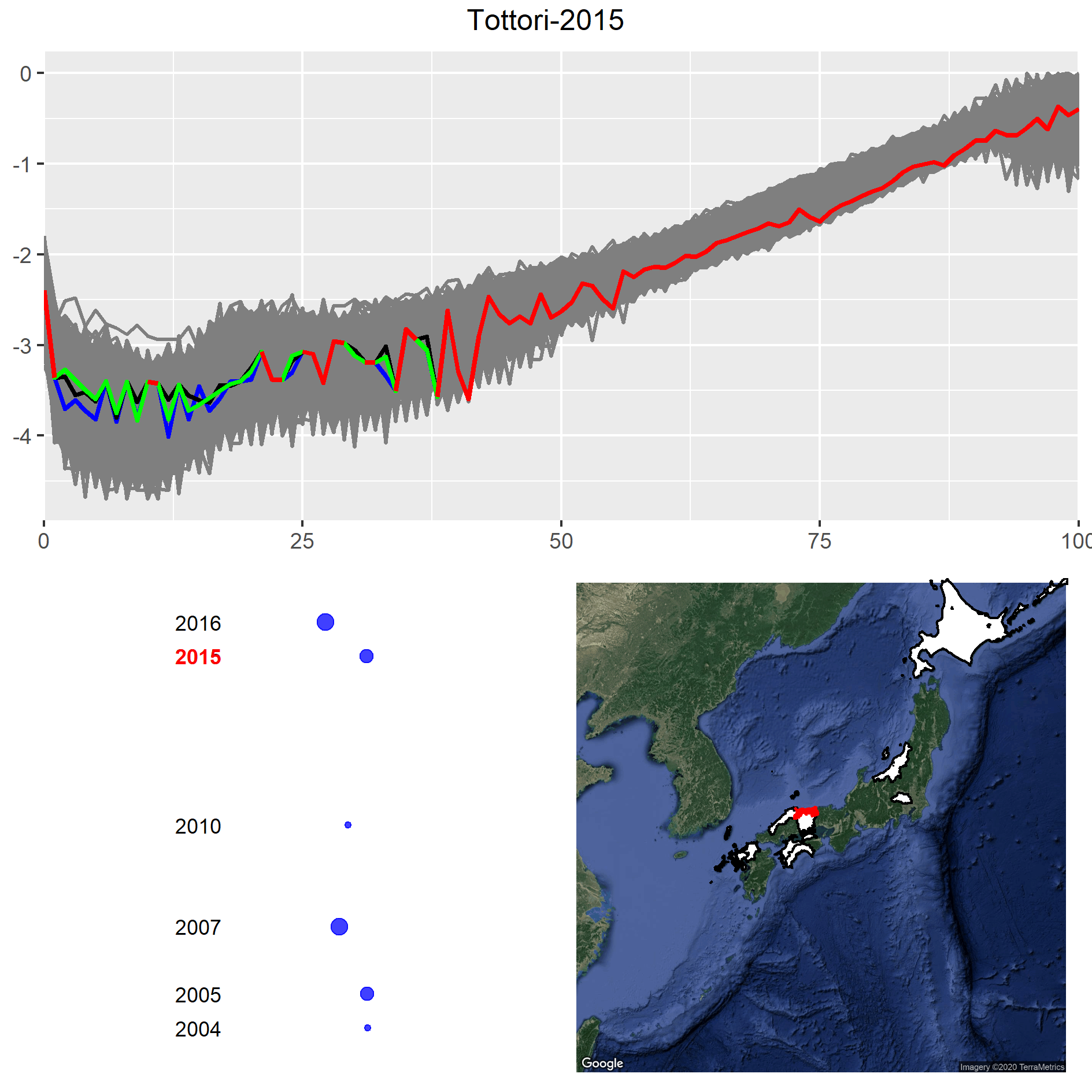}
\caption{\small{Real partially observed function of the sample observed in $6$ fragments, Tottori-2015. The top panel presents the reconstruction given by Reg. Regression method by \cite{Kraus2015} (black), Opt. Operator method by \cite{lieblReco2019} (green) and the depth-based method (blue). The bottom panel shows the year of the most important functions of the envelope, and the right map shows its prefectures.}}\label{fig:JAPANother}
\end{center}
\end{figure}

In this case study, the poor availability of complete functions invalidates the possibility of doing a resampling exercise like the one we have done for the AEMET data set. Then, we are only able to illustrate some real situations. Figures~\ref{fig:JAPANshortest} and~\ref{fig:JAPANother} present two real reconstruction problems and the results obtained from the three methods. In Figure~\ref{fig:JAPANshortest}, we reconstruct the shortest available curve, ``Saitama-2007", that was only available in a very short interval of mortality rates for the youngest cohorts. Reg. Regression \citep{Kraus2015} and Opt. Operator \citep{lieblReco2019} produce smooth results (black and green respectively). The depth-based method produces more spiky results in concordance with other available curves. The bottom panel presents the bubble plot illustrating the period of the envelope functions and the prefectures in the map. Figure~\ref{fig:JAPANother} presents a case with the function ``Tottori-2015" that is not observed in six intervals fragments (domain where only the red curve is visible). 

\section{Conclusion}\label{sec:conclu}

This article introduces a non-parametric method to reconstruct samples of incomplete functional data. Our proposal relies on the concept of depth for POFD to select a subset of sample curves that is defined to share shape and magnitude with the observed part of the curve to predict. 
We term this subset of sample curves as envelope, and we use them to proposed a point reconstruction method based on a weighted average. These weights only depend on a parameter we set by minimizing the MSE where the curve to predict is observed.  

We compare the performance of the new method with other alternatives of the literature. Our simulation exercises consider simulated and real data and various random procedures to generate incomplete data scenarios. The available reconstruction methods seem to be unbeatable in our settings when the covariance can be efficiently estimated. These favorable circumstances are exemplified with Gaussian processes with stationary covariance functions and other more complex covariance regimes, including a considerable proportion of completely observed curves. In contrast, our method outperforms when the covariance can not be properly estimated due to a richer covariance structure and highly scarce data settings. To show that, we test the methods under severe incomplete data settings and introduce more complex covariance structures. Particularly, we decrease the number of completely observed functions up to zero and consider real data with complex covariance structures such as yearly age-specific mortality and yearly temperature data. Finally, our simulation exercises show that our proposal can provide a reasonable reconstruction output when every function is partially observed or, in other words, when there are no complete functions in the sample. 

The depth-based method requires the Missing-Completely-at-Random assumption, as it is standard in the literature. This assumption implies that the partially observed functions cover densely the reconstruction domain and that the observability process is not conditional to external information. Future research could allow for specific relationships between the functional process and the process that generates partial observability. In addition, the depth-based algorithm requires a notion of proximity between POFD that we fill with a $L_2$ distance between the observed segments. Developments along this line would also be valuable for our proposal.

In summary, this article provides an alternative data-driven and model-free method to reconstruct partially observed functional data preferable under challenging scenarios like the ones presented here. Last but not least, we believe that the interpretability of the results might be helpful to provide different insides into the data under analysis.

\section*{Acknowledgments}

Antonio El\'ias was supported by the Ministerio de Educaci\'on, Cultura y Deporte under grant FPU15/00625 and the research stay grant EST17/00841. Antonio Elías and Ra\'ul Jim\'enez were partially supported by the Spanish Ministerio de Econom\'ia y Competitividad under grant ECO2015-66593-P. Part of this article was conducted during a stay at Australian National University. Antonio Elías is grateful to Han Lin Shang for his hospitality and insightful and constructive discussions.

\newpage
\bibliographystyle{apalike}
\bibliography{ref}

\end{document}